\documentstyle[12pt]{article}

\font\twlgot =eufm10 scaled \magstep1 \font\egtgot =eufm8
\font\sevgot =eufm7 \font\twlmsb =msbm10 scaled \magstep1
\font\egtmsb =msbm8 \font\sevmsb =msbm7

\newfam\gotfam
\def\pgot{\fam\gotfam\twlgot}
\textfont\gotfam\twlgot \scriptfont\gotfam\egtgot
\scriptscriptfont\gotfam\sevgot
\def\got{\protect\pgot}
\newfam\msbfam
\textfont\msbfam\twlmsb \scriptfont\msbfam\egtmsb
\scriptscriptfont\msbfam\sevmsb
\def\Bbb{\protect\pBbb}
\def\pBbb{\relax\ifmmode\expandafter\Bb\else\typeout{You cann't use
Bbb in text mode}\fi}
\def\Bb #1{{\fam\msbfam\relax#1}}

\newcommand{\ccG}{{\got g}}

\def\thebibliography#1{\section*{References}\list
  {[\arabic{enumi}]}{\settowidth\labelwidth{#1}\leftmargin\labelwidth
    \advance\leftmargin\labelsep
    \usecounter{enumi}}
    \def\newblock{\hskip .11em plus .33em minus .07em}
    \sloppy\clubpenalty4000\widowpenalty4000
    \sfcode`\.=1000\relax}

\def\op#1{\mathop{\fam0 #1}\limits}

\newcommand{\id}{{\rm Id\,}}

\newcommand{\beq}{\begin{equation}}
\newcommand{\eeq}{\end{equation}}
\newcommand{\ben}{\begin{eqnarray}}
\newcommand{\een}{\end{eqnarray}}
\newcommand{\be}{\begin{eqnarray*}}
\newcommand{\ee}{\end{eqnarray*}}
\newcommand{\bea}{\begin{eqalph}}
\newcommand{\eea}{\end{eqalph}}

\newcommand{\cT}{{\cal T}}

\newcommand{\cD}{{\cal D}}
\newcommand{\cR}{{\cal R}}

\newcommand{\cL}{{\cal L}}
\newcommand{\cE}{{\cal E}}

\newcommand{\cF}{{\cal F}}

\newcommand{\cS}{{\cal S}}

\newcommand{\cJ}{{\cal J}}
\newcommand{\bL}{{\bf L}}

\newcommand{\bF}{{\bf F}}

\newcommand{\rL}{{\rm L}}
\newcommand{\rs}{{\rm s}}

\newcommand{\al}{\alpha}

\newcommand{\bt}{\beta}
\newcommand{\dl}{\delta}
\newcommand{\la}{\lambda}

\newcommand{\om}{\omega}

\newcommand{\m}{\mu}
\newcommand{\n}{\nu}

\newcommand{\g}{\gamma}
\newcommand{\G}{\Gamma}
\newcommand{\th}{\theta}

\newcommand{\si}{\sigma}
\newcommand{\Si}{\Sigma}
\newcommand{\w}{\wedge}
\newcommand{\wt}{\widetilde}
\newcommand{\wh}{\widehat}
\newcommand{\ol}{\overline}
\newcommand{\dr}{\partial}
\newcommand{\ar}{\op\longrightarrow}
\newcommand{\ot}{\otimes}
\newcommand{\ap}{\approx}
\newcommand{\ve}{\varepsilon}

\newcounter{theorem}
\newcounter{remark}
\newcounter{proposition}
\newcounter{lemma}
\newcounter{corollary}
\newcounter{definition}

\def\theremark{\arabic{remark}}

\def\thedefinition{\arabic{definition}}

\newenvironment{proof}{\noindent
{\bf Proof.}}{\hfill $\Box$ \medskip}
\newenvironment{rem}{\refstepcounter{remark}\medskip\noindent{\bf Remark
\theremark.}}{\medskip}

\newenvironment{theo}{\refstepcounter{definition} \medskip
\noindent{\bf Theorem \thedefinition.}}{\medskip}

\newenvironment{cor}{\refstepcounter{definition} \medskip
\noindent{\bf Corollary \thedefinition.}}{\medskip}

\newcommand{\mar}[1]{}

\begin{document}
\hbox{}

{\parindent=0pt

\begin{center}

{\large\bf CLASSICAL GAUGE GRAVITATION THEORY}
\bigskip

{\sc G.SARDANASHVILY}

{\it Department of Theoretical Physics, Moscow State University}

\end{center}

\bigskip
\bigskip
{\small Classical gravitation theory is formulated as gauge theory
on natural bundles where gauge symmetries are general covariant
transformations and a gravitational field is a Higgs field
responsible for their spontaneous symmetry breaking.


} }

\section{Introduction}

Classical field theory admits a comprehensive mathematical
formulation in the geometric terms of smooth fibre bundles
\cite{book09,sard08}. For instance, Yang--Mills gauge theory is
theory of principal connections on principal bundles.

Gauge gravitation theory as particular classical field theory also
is formulated in the terms of fibre bundles.

\begin{rem}
A first gauge model of gravity was suggested by Utiyama \cite{uti}
in 1956 just two years after the birth of gauge theory itself. He
was first who generalized the original gauge model of Yang and
Mills for $SU(2)$ to an arbitrary symmetry Lie group and, in
particular, to a Lorentz group in order to describe gravity.
However, he met the problem of treating general covariant
transformations and a pseudo-Riemannian metric which had no
partner in Yang--Mills gauge theory. To eliminate this drawback,
representing a gravitational field as a gauge field of a
translation group was attempted (see \cite{heh,iva} for a review).
Since the Poincar\'e group comes from the Wigner--In\"onii
contraction of de Sitter groups $SO(2,3)$ and $SO(1,4)$ and it is
a subgroup of a conformal group, gauge theories on fibre bundles
$Y\to X$ with these structure groups were also considered
\cite{ivan,tseytl}. In a different way, gravitation theory was
formulated as gauge theory with a Lorentz reduced structure where
a metric gravitational field was treated as the corresponding
Higgs field \cite{iva,sardz}.
\end{rem}

Studying gauge gravitation theory, we believe reasonable  to
require that it incorporates Einstein's General Relativity and,
therefore, it should be based on the relativity and equivalence
principles reformulated in the fibre bundle terms.

In these terms, the relativity principle states that gauge
symmetries of classical gravitation theory are general covariant
transformations. Let us emphasized that these gauge symmetries
differ from gauge symmetries of the above mentioned Yang--Mills
gauge theory which are vertical automorphisms of principal
bundles. Fibre bundles possessing general covariant
transformations constitute the category of so called natural
bundles \cite{book09,kol}.

Let $\pi:Y\to X$ be a smooth fibre bundle. Any automorphism
$(\Phi,f)$ of $Y$, by definition, is projected as $\pi \circ \Phi=
f\circ \pi$ onto a diffeomorphism $f$ of its base $X$. The
converse is not true. A fibre bundle $Y\to X$ is called the
natural bundle if there exists a monomorphism
\be
{\rm Diff} X\ni f\to\wt f\in {\rm Aut} Y
\ee
of the group of diffeomorphisms of $X$ to the group of bundle
automorphisms of $Y\to X$. Automorphisms $\wt f$ are called
general covariant transformations of $Y$. The group of
automorphisms of a natural bundle is a semi-direct product of a
subgroup of vertical automorphisms of $Y$ (over $\id X$) and a
subgroup of general covariant transformations.

Accordingly, there exists the functorial lift of any vector field
$\tau$ on $X$ to a vector field $\ol\tau$ on $Y$ such that
$\tau\mapsto\ol\tau$ is a monomorphism of the Lie algebra $\cT(X)$
of vector field on $X$ to that $\cT(T)$ of vector fields on $Y$.
This functorial lift $\ol\tau$ is an infinitesimal generator of a
local one-parameter group of local general covariant
transformations of $Y$.

The tangent bundle $TX$ of $X$ exemplifies a natural bundle. Any
diffeomorphism $f$ of $X$ gives rise to the tangent automorphisms
$\wt f=Tf$ of $TX$ which is a general covariant transformation of
$TX$. The tangent bundle possess a structure group
\mar{gl4}\beq
GL_4=GL^+(4,\Bbb R). \label{gl4}
\eeq
The associated principal bundle is a fibre bundle $LX$ of frames
in the tangent spaces to $X$. It also is a natural bundle.
Moreover, all fibre bundles associated with $LX$ are natural
bundles. Principal connections on $LX$ yield linear connections on
the tangent bundle $TX$ and other associated bundles. They are
called the world connections.

Following the relativity principle, one thus should develop
gravitation theory as a gauge theory of principal connections on a
principal frame bundle $LX$ over an oriented four-dimensional
smooth manifold $X$, called the world manifold $X$
\cite{book09,sard06}.

The equivalence principle reformulated in geometric terms requires
that the structure group $GL_4$ (\ref{gl4}) of a frame bundle $LX$
and associated bundles is reducible to a Lorentz group $SO(1,3)$
\cite{iva,sardz}. It means that these fibre bundles admit atlases
with $SO(1,3)$-valued transition functions or, equivalently, that
there exist principal subbubdles of $LX$ with a Lorentz structure
group. This is the case of spontaneous symmetry breaking.

Spontaneous symmetry breaking in classical gauge theory on a
principal bundle $P\to X$ with a structure Lie group $G$ is
characterized as a reduction of this structure group to its closed
(consequently, Lie) subgroup $H$ \cite{book09,sard92,higgs}. By
virtue of the well-known theorem \cite{book09,kob}, there is
one-to-one correspondence between the $H$-principal subbundles
$P^h$ of $P$ and the global sections $h$ of the quotient bundle
$P/H\to X$ which are treated as classical Higgs fields.

Accordingly, in gauge gravitation theory based on the equivalence
principle, there is one-to-one correspondence between the Lorentz
principal subbundles of a frame bundle $LX$ (called the Lorentz
reduced structures) and the global sections of the quotient bundle
\mar{b3203}\beq
\Si_{\rm PR}= LX/SO(1,3),\label{b3203}
\eeq
which are pseudo-Riemannian metrics on a world manifold. In
Einstein's General Relativity, they are identified with
gravitational fields.

Thus, we come to gauge gravitation theory as metric-affine
gravitation theory whose dynamic variables are world connections
and pseudo-Riemannian metrics on a world manifold $X$ (Section 6).
They are treated as gauge fields and Higgs fields, respectively.

There is the extensive literature on metric-affine gravitation
theory \cite{heh,heh07,obukh}. However, one often formulates it as
gauge theory of affine connections, that is wrong (Section 11).

The character of gravity as a Higgs field responsible for
spontaneous breaking of general covariant transformations is
displayed as follows. Given different gravitational fields, the
representations (\ref{L4'}) of holonomic coframes $\{dx^\m\}$ by
Dirac matrices acting on Dirac spinor fields are nonequivalent
(Section 10). Consequently, Dirac operators in the presence of
different gravitational fields fails to be equivalent, too. In
particular, it follows that a Dirac spinor field can be considered
only in a pair with a certain gravitational field. A total system
of such pairs is described by sections of the composite bundle
$S\to X$ (\ref{qqz}), where $S\to \Si_{\rm T}$ is a spinor bundle,
whereas $S\to X$ is a natural bundle.

\begin{rem}
Since the Dirac operators in the presence of different
gravitational fields are nonequivalent, Dirac spinor fields fail
to be considered, e.g., in the case of a superposition of
different gravitational fields. Therefore, quantization of a
metric gravitational field fails to satisfy the superposition
principle, and one can suppose that a metric gravitational field
as a Higgs field is non-quantized in principle.
\end{rem}

Being reduced to a Lorentz group, a structure group of a frame
bundle $LX$ also is reduced to a maximal compact subgroup $SO(3)$
of $SO(1,3)$. The associated Higgs field is a spatial distribution
which defines a space-time structure on a world manifold $X$
(Section 5).

Since general covariant transformations are symmetries of a
metric-affine gravitation Lagrangian, the corresponding
conservation law holds (Section 7). It is an energy-momentum
conservation law.  Because general covariant transformations are
gauge transformations depending on derivatives of gauge
parameters, the corresponding energy-momentum current reduces to a
superpotential \cite{book09,sard09}. This is a generalized Komar
superpotential \cite{giacqg,sard97}.

Since general covariant transformations are gauge symmetries of a
gravitation Lagrangian $L_{\rm MA}$,  the Euler--Lagrange operator
$\dl L_{\rm MA}$ (\ref{eeu}) of this Lagrangian obeys the complete
Noether identities (\ref{333}). Because they are irreducible, one
obtains the BRST extension (\ref{444}) of general covariant
transformations and that (\ref{555}) of an original metric-affine
Lagrangian. This is a necessary step towards quantization of
classical gauge gravitation theory \cite{gauge05,book09}.

\section{Natural bundles}

Let $\pi:Y\to X$ be a smooth fibre bundle coordinated by $(x^\la,
y^i)$. Given a one-parameter group $(\Phi_t,f_t)$ of automorphisms
of $Y$, its infinitesimal generator is a projectable vector field
\mar{klo}\beq
u=\tau^\la(x^\m)\dr_\la + u^i(x^\m,y^j)\dr_i \label{klo}
\eeq
on $Y$ which is projected onto a vector field
$\tau=\tau^\la\dr_\la$ on $X$, whose flow is a one-parameter group
$(f_t)$ of diffeomorphisms of $X$. Conversely, let
$\tau=\tau^\la\dr_\la$ be a vector field on $X$. Its lift to some
projectable vector field (\ref{klo}) on $Y$ always exists. For
instance, given a connection
\be
\G=dx^\la\ot(\dr_\la +\G^i_\la(x^\m,y^j)\dr_i)
\ee
on $Y\to X$, a vector field $\tau$ on $X$ gives rise to a
horizontal vector field
\be
\G\tau=\tau\rfloor\G=\tau_\la (\dr_\la+\G^i_\la\dr_i)
\ee
on $Y$. The horizontal lift $\tau\to\G\tau$ yields a monomorphism
of a $C^\infty(X)$-module $\cT(X)$ of vector fields on $X$ to a
$C^\infty(Y)$-module $\cT(Y)$ of vector fields on $Y$, but this
monomorphism is not a Lie algebra morphism, unless $\G$ is a flat
connection.

We address the category of natural bundles $Y\to X$
\cite{book09,kol} admitting the functorial lift $\wt\tau$ onto $Y$
of any vector field $\tau$ on $X$ such that $\tau\to\ol\tau$ is a
Lie algebra monomorphism
\be
\cT(X)\to \cT(T), \qquad [\wt\tau,\wt\tau']=\wt{[\tau,\tau']}.
\ee
This functorial lift $\wt\tau$, by definition, is an infinitesimal
generator of a local one-parameter group of general covariant
transformations of $Y$.

Natural bundles are exemplified by tensor products
\mar{sp20}\beq
T=(\op\ot^m TX)\ot(\op\ot^k T^*X) \label{sp20}
\eeq
of the tangent $TX$ and cotangent $T^*X$ bundles of $X$ as
follows. Given a manifold atlas $\{(U_i,\phi_i)\}$ of $X$, the
tangent bundle $\pi_X:TX\to X$ admits a holonomic bundle atlas
\mar{nn2}\beq
\{(U_i, T\phi_i:\pi_X^{-1}(U_i)\to U_i\times \Bbb R^4)\},
\label{nn2}
\eeq
where $T\phi_i$ is the tangent morphism to $\phi_i$. With this
atlas, $TX$ is provided with holonomic bundle coordinates
\be
(x^\m,\dot x^\m), \qquad \dot x'^\m =\frac{\dr x'^\m}{\dr
x^\nu}\dot x^\nu,
\ee
where $(\dot x^\m)$ are fibre coordinates with respect to
holonomic frames $\{\dr_\m\}$. Accordingly, the tensor bundle
(\ref{sp20}) is endowed with holonomic bundle coordinates
$(x^\m,x^{\al_1\cdots\al_m}_{\bt_1\cdots\bt_k})$, where
\be
(x^\m,\dot x_\m), \qquad \dot x'_\m =\frac{\dr x^\nu}{\dr
x_\m}\dot x_\nu,
\ee
are holonomic bundle coordinates on the cotangent bundle $T^*X$ of
$X$. Then given a vector field $\tau$ on $X$, its functorial lift
onto the tensor bundle (\ref{sp20}) takes the form
\be \wt\tau =
\tau^\m\dr_\m + [\dr_\nu\tau^{\al_1}\dot
x^{\nu\al_2\cdots\al_m}_{\bt_1\cdots\bt_k} + \ldots
-\dr_{\bt_1}\tau^\nu \dot
x^{\al_1\cdots\al_m}_{\nu\bt_2\cdots\bt_k} -\ldots]\dot \dr
_{\al_1\cdots\al_m}^{\bt_1\cdots\bt_k}, \qquad \dot\dr_\la =
\frac{\dr}{\dr\dot x^\la}.
\ee

Tensor bundles over a world manifold $X$ have the structure group
$GL_4$ (\ref{gl4}). An associated principal bundle is the above
mentioned frame bundle $LX$. Its (local) sections are called frame
fields. Given the holonomic atlas (\ref{nn2}) of the tangent
bundle $TX$, every element $\{H_a\}$ of a frame bundle $LX$ takes
the form $H_a=H^\m_a\dr_\m$, where $H^\m_a$ is a matrix of the
natural representation of a group $GL_4$ in $\Bbb R^4$. These
matrices constitute bundle coordinates
\be
(x^\la, H^\m_a), \qquad H'^\m_a=\frac{\dr x'^\m}{\dr
x^\la}H^\la_a,
\ee
on $LX$ associated to its holonomic atlas
\mar{tty}\beq
\Psi_T=\{(U_\iota, z_\iota=\{\dr_\m\})\}, \label{tty}
\eeq
given by local frame fields $z_\iota=\{\dr_\m\}$. With respect to
these coordinates, the canonical right action of $GL_4$ on $LX$
reads $GL_4\ni g: H^\m_a\to H^\m_bg^b{}_a$.

A frame bundle $LX$ is equipped with a canonical $\Bbb R^4$-valued
one-form
\mar{b3133'}\beq
\th_{LX} = H^a_\m dx^\m\ot t_a,\label{b3133'}
\eeq
where $\{t_a\}$ is a fixed basis for $\Bbb R^4$ and $H^a_\m$ is
the inverse matrix of $H^\m_a$.

A frame bundle $LX\to X$ is natural. Any diffeomorphism $f$ of $X$
gives rise to a principal automorphism
\mar{025}\beq
\wt f: (x^\la, H^\la_a)\to (f^\la(x),\dr_\m f^\la H^\m_a)
\label{025}
\eeq
of $LX$ which is its general covariant transformation. Given a
(local) one-parameter group of diffeomorphisms of $X$ and its
infinitesimal generator $\tau$, the lift (\ref{025}) yields a
functorial lift
\be
\wt\tau=\tau^\m\dr_\m +\dr_\nu\tau^\al H^\nu_a\frac{\dr}{\dr
H^\al_a}
\ee
onto $LX$ of a vector field $\tau$ on $X$ which is defined by the
condition $\bL_{\wt\tau}\th_{LX}=0$.

Let $Y=(LX\times V)/GL_4$ be an $LX$-associated bundle with a
typical fibre $V$. It admits a lift of any diffeomorphism $f$ of
its base to an automorphism
\be
f_Y(Y)=(\wt f(LX)\times V)/GL_4
\ee
of $Y$ associated with the principal automorphism $\wt f$
(\ref{025}) of a frame bundle $LX$. Thus, all bundles associated
with a frame bundle $LX$ are natural bundles.

\section{World connections}

Let $TX$ be the tangent bundle of a world manifold $X$. With
respect to holonomic coordinates $(x^\la,\dot x^\la)$, a linear
connection on $TX$ takes the form
\mar{B}\beq
\G= dx^\la\otimes (\dr_\la +\G_\la{}^\m{}_\n \dot x^\n
\dot\dr_\m). \label{B}
\eeq
Since $TX$ is associated with a frame bundle $LX$, every linear
connection (\ref{B}) is associated with a principal connection on
$LX$. We agree to call them world connections.

A curvature of a world connection is defined as that of the
connection (\ref{B}). It reads
\mar{1203}\ben
&& R=\frac12R_{\la\m}{}^\al{}_\bt\dot x^\bt dx^\la\w dx^\m\ot\dot\dr_\al,
\label{1203}\\
&& R_{\la\m}{}^\al{}_\bt = \dr_\la \G_\m{}^\al{}_\bt - \dr_\m
\G_\la{}^\al{}_\bt + \G_\la{}^\g{}_\bt \G_\m{}^\al{}_\g -
\G_\m{}^\g{}_\bt \G_\la{}^\al{}_\g. \nonumber
\een
Due to the canonical splitting of the vertical tangent bundle
\mar{mos163}\beq
VTX=TX\times TX \label{mos163}
\eeq
of $TX$, the curvature $R$ (\ref{1203}) can be represented by a
tangent-valued two-form
\mar{1203a}\beq
R=\frac12R_{\la\m}{}^\al{}_\bt\dot x^\bt dx^\la\w dx^\m\ot\dr_\al
\label{1203a}
\eeq
on $TX$. Due to this representation, the Ricci tensor
\mar{ric}\beq
R_c=\frac12R_{\la\m}{}^\la{}_\bt dx^\m\ot dx^\bt \label{ric}
\eeq
of a world connection $\G$ is defined.

By a torsion of a world connection is meant that of the connection
$\G$ (\ref{B}) on the tangent bundle $TX$ with respect to the
canonical soldering form
\mar{z117'}\beq
\th_J=dx^\m\ot\dot\dr_\m \label{z117'}
\eeq
on $TX$. It reads
\mar{191}\beq
T =\frac12 T_\m{}^\n{}_\la  dx^\la\w dx^\m\ot \dot\dr_\n, \qquad
T_\m{}^\n{}_\la  = \G_\m{}^\n{}_\la - \G_\la{}^\n{}_\m.
\label{191}
\eeq
A world connection is said to be symmetric if its torsion
(\ref{191}) vanishes, i.e., $\G_\m{}^\n{}_\la = \G_\la{}^\n{}_\m$.
Owing to the vertical splitting (\ref{mos163}), the torsion form
$T$ (\ref{191}) of $\G$ can be written as a tangent-valued
two-form
\mar{mos164}\beq
T =\frac12 T_\m{}^\n{}_\la  dx^\la\w dx^\m\ot \dr_\n
\label{mos164}
\eeq
on $X$. One also introduces a soldering torsion form
\mar{mos160}\beq
T=T_\m{}^\n{}_\la \dot x^\la dx^\m\ot\dot \dr_\nu. \label{mos160}
\eeq

Given a world connection $\G$ (\ref{B}) and its soldering torsion
form $T$ (\ref{mos160}), the sum $\G+c T$, $c\in\Bbb R$, is a
world connection. In particular, every world connection $\G$
defines a unique symmetric world connection $\G'=\G -T/2$.

Being associated with a principal connection on $LX$, a world
connection is represented by a section of the quotient bundle
\mar{015}\beq
C_{\rm W}=J^1LX/GL_4\to X, \label{015}
\eeq
where $J^1LX$ is the first order jet manifold of sections of
$LX\to X$ \cite{book09}. We agree to call $C_{\rm W}$ (\ref{015})
the bundle of world connections. With respect to the holonomic
atlas $\Psi_T$ (\ref{tty}), it is provided with the bundle
coordinates
\be
(x^\la, k_\la{}^\nu{}_\al), \qquad k'_\la{}^\nu{}_\al=
\left[\frac{\dr x'^\nu}{\dr x^\g} \frac{\dr x^\bt}{\dr x'^\al}
k_\m{}^\g{}_\bt + \frac{\dr x^\bt}{\dr x'^\al}\frac{\dr^2
x'^\nu}{\dr x^\m\dr x^\bt} \right] \frac{\dr x^\m}{\dr x'^\la},
\ee
so that, for any section $\G$ of $C_{\rm W}\to X$, its coordinates
$k_\la{}^\nu{}_\al\circ \G=\G_\la{}^\nu{}_\al$ are components of
the world connection $\G$ (\ref{B}).

Though the bundle of world connections $C_{\rm W}\to X$
(\ref{015}) is not $LX$-associated, it is a natural bundle. It
admits a functorial lift
\be
 \wt\tau_C = \tau^\m\dr_\m
+[\dr_\nu\tau^\al k_\m{}^\nu{}_\bt - \dr_\bt\tau^\nu
k_\m{}^\al{}_\nu - \dr_\m\tau^\nu k_\nu{}^\al{}_\bt +
\dr_{\m\bt}\tau^\al]\frac{\dr}{\dr k_\m{}^\al{}_\bt}
\ee
of any vector field $\tau$ on $X$.

The first order jet manifold $J^1C_{\rm W}$ of a bundle of world
connections possesses the canonical splitting
\mar{0101}\ben
&&k_{\la\m}{}^\al{}_\bt =
 \frac12(k_{\la\m}{}^\al{}_\bt - k_{\m\la}{}^\al{}_\bt +
k_\la{}^\g{}_\bt k_\m{}^\al{}_\g -k_\m{}^\g{}_\bt
k_\la{}^\al{}_\g) + \label{0101}\\
&& \qquad \frac12(k_{\la\m}{}^\al{}_\bt +
k_{\m\la}{}^\al{}_\bt - k_\la{}^\g{}_\bt k_\m{}^\al{}_\g +
k_\m{}^\g{}_\bt k_\la{}^\al{}_\g)=\frac12(\cR_{\la\m}{}^\al{}_\bt
+\cS_{\la\m}{}^\al{}_\bt), \nonumber
\een
so that, if $\G$ is a section of $C_{\rm W}\to X$, then
$\cR_{\la\m}{}^\al{}_\bt\circ J^1\G=R_{\la\m}{}^\al{}_\bt$ are
components of the curvature (\ref{1203}).

A world manifold $X$ is called flat if it admits a flat world
connection $\G$. By virtue of the well-known theorem, there exists
a bundle atlas of $TX$ with constant transition functions such
that $\G=dx^\la\ot\dr_\la$ relative to this atlas \cite{book09}.
However, such an atlas is not holonomic in general. Therefore, the
torsion form $T$ (\ref{191}) of a flat connection $\G$ need not
vanish.

A world manifold $X$ is called parallelizable if the tangent
bundle $TX\to X$ is trivial. A parallelizable manifold is flat. A
flat manifold is parallelizable if it is simply connected.

\section{Lorentz reduced structure}

As was mentioned above, gravitation theory on a world manifold $X$
is classical field theory with spontaneous symmetry breaking
described by Lorentz reduced structures of a frame bundle $LX$. We
deal with the following Lorentz and proper Lorentz reduced
structures.

By a Lorentz reduced structure is meant a reduced principal
$SO(1,3)$-subbundle $L^gX$, called the Lorentz subbundle, of a
frame bundle $LX$.

Let L$=SO^0(1,3)$ be a proper Lorentz group. Recall that
$SO(1,3)=\Bbb Z_2\times$L. A proper Lorentz reduced structure is
defined as a reduced L-subbundle $L^hX$ of $LX$.

If a world manifold $X$ is simply connected, there is one-to-ne
correspondence between the Lorentz and proper Lorentz reduced
structures.

One can show that different proper Lorentz subbundles $L^hX$ and
$L^{h'}X$ of a frame bundle $LX$ are isomorphic as principal
L-bundles. This means that there exists a vertical automorphism of
a frame bundle $LX$ which sends $L^hX$ onto $L^{h'}X$
\cite{book09,higgs}. If a world manifold $X$ is simply connected,
the similar property of Lorentz subbundles also is true.

There is the well-known topological obstruction to the existence
of a Lorentz structure on a world manifold $X$. All non-compact
manifolds and compact manifolds whose Euler characteristic equals
zero admit a Lorentz reduced structure \cite{dods}.

By virtue of the above mentioned theorem, there is one-to-one
correspondence between the principal L-subbundles $L^hX$ of a
frame bundle $LX$ and the global sections $h$ of a quotient bundle
\mar{5.15}\beq
\Si_{\rm T}=LX/\rL,  \label{5.15}
\eeq
called the tetrad bundle. This is an $LX$-associated bundle with
the typical fibre $GL_4/$L. Its global sections are called the
tetrad fields. The fibre bundle (\ref{5.15}) is a two-fold
covering $\zeta: \Si_{\rm T}\to \Si_{\rm PR}$ of the metric bundle
$\Si_{\rm PR}$ (\ref{b3203}). In particular, every tetrad field
$h$ defines a unique pseudo-Riemannian metric $g=\zeta\circ h$.
For the sake of convenience, one usually identifies a metric
bundle with an open subbundle of the tensor bundle $\Si_{\rm
PR}\subset \op\vee^2 TX$. Therefore, the metric bundle $\Si_{\rm
PR}$ (\ref{b3203}) can be equipped with bundle coordinates
$(x^\la, \si^{\m\nu})$.

Every tetrad field $h$ defines an associated Lorentz bundle atlas
\mar{lat}\beq
\Psi^h=\{(U_\iota,z_\iota^h=\{h_a\})\} \label{lat}
\eeq
of a frame bundle $LX$ such that the corresponding local sections
$z_\iota^h$ of $LX$ take their values into a proper Lorentz
subbundle $L^hX$ and the transition functions of $\Psi^h$
(\ref{lat}) between the frames $\{h_a\}$ are L-valued. The frames
(\ref{lat}):
\mar{b3211a}\beq
\{h_a =h_a^\m(x)\dr_\m\}, \qquad h_a^\m=H_a^\m\circ z_\iota^h,
\qquad x\in U_\iota, \label{b3211a}
\eeq
are called the tetrad frames.

Given a Lorentz bundle atlas $\Psi^h$, the pull-back
\mar{b3211}\beq
h=h^a\ot t_a=z_\iota^{h*}\th_{LX}=h_\la^a(x) dx^\la\ot t_a
\label{b3211}
\eeq
of the canonical form $\th_{LX}$ (\ref{b3133'}) by a local section
$z_\iota^h$ is called the (local) tetrad form. It determines
tetrad coframes
\mar{b3211'}\beq
\{h^a =h^a_\m(x)dx^\m\}, \qquad x\in U_\iota, \label{b3211'}
\eeq
in the cotangent bundle $T^*X$. They are the dual of the tetrad
frames (\ref{b3211a}). The coefficients $h_a^\m$ and $h^a_\m$ of
the tetrad frames (\ref{b3211a}) and coframes (\ref{b3211'}) are
called the tetrad functions. They are transition functions between
the holonomic atlas $\Psi_T$ (\ref{tty}) and the Lorentz atlas
$\Psi^h$ (\ref{lat}) of a frame bundle $LX$.

With respect to the Lorentz atlas $\Psi^h$ (\ref{lat}), a tetrad
field $h$ can be represented by the $\Bbb R^4$-valued tetrad form
(\ref{b3211}). Relative to this atlas, the corresponding
pseudo-Riemannian metric $g=\zeta\circ h$ takes the well-known
form
\mar{mos175}\beq
g=\eta(h\ot h)=\eta_{ab}h^a\ot h^b, \qquad
g_{\m\nu}=h_\m^ah_\nu^b\eta_{ab}, \label{mos175}
\eeq
where $\eta={\rm diag}(1,-1,-1,-1)$ is the Minkowski metric in
$\Bbb R^4$ written with respect to its fixed basis $\{t_a\}$. It
is readily observed that the tetrad coframes $\{h^a\}$
(\ref{b3211'}) and the tetrad frames $\{h_a\}$ (\ref{b3211a}) are
orthornormal relative to the pseudo-Riemannian metric
(\ref{mos175}), namely:
\be
g^{\m\nu}h^a_\mu h^b_\nu=\eta^{ab}, \qquad g_{\m\nu}h_a^\mu
h_b^\nu=\eta_{ab}.
\ee
Therefore, their components $h^0$, $h_0$ and $h^i$, $h_i$,
$i=1,2,3$, are called time-like and spatial, respectively.

Given a pseudo-Riemannian metric $g$, any world connection $\G$
(\ref{B}) admits a splitting
\mar{mos191}\beq
\G_{\m\n\al}=\{_{\m\n\al}\} +S_{\m\n\al} +\frac12 C_{\m\n\al}
\label{mos191}
\eeq
in the Christoffel symbols
\mar{b1.400}\beq
\{_{\m\n\al}\}= -\frac12(\dr_\m g_{\nu\al} + \dr_\al
g_{\nu\m}-\dr_\nu g_{\m\al}), \label{b1.400}
\eeq
the non-metricity tensor
\mar{mos193}\beq
C_{\m\n\al}=C_{\m\al\n}=\nabla^\G_\m g_{\n\al}=\dr_\m g_{\n\al}
+\G_{\m\n\al} + \G_{\m\al\n} \label{mos193}
\eeq
and the contorsion
\mar{mos202}\beq
S_{\m\n\al}=-S_{\m\al\n}=\frac12(T_{\n\m\al} +T_{\n\al\m} +
T_{\m\n\al}+ C_{\al\n\m} -C_{\n\al\m}), \label{mos202}
\eeq
where $T_{\m\nu\al}=-T_{\al\nu\m}$ are coefficients of the torsion
form (\ref{mos164}) of $\G$.

A world connection $\G$ is called a metric connection for a
pseudo-Riemannian metric $g$ if $g$ is its integral section, i.e.,
the metricity condition
\mar{mos203}\beq
\nabla^\G_\m g_{\n\al}=0 \label{mos203}
\eeq
holds. A metric connection reads
\mar{mos204}\beq
\G_{\m\n\al}=\{_{\m\n\al}\} + \frac12(T_{\n\m\al} +T_{\n\al\m} +
T_{\m\n\al}). \label{mos204}
\eeq
For instance, the Levi--Civita connection (\ref{b1.400}) is a
torsion-free metric connection.

A principal connection on a proper Lorentz subbundle $L^hX$ of a
frame bundle $LX$ is called the Lorentz connection. By virtue of
the well known theorem \cite{book09,kob}, this connection is
extended to a principal connection $\G$ on a frame bundle $LX$. It
also is called the Lorentz connection. The associated linear
connection (\ref{B}) on the tangent bundle $TX$ with respect to a
Lorentz atlas $\Psi^h$ reads
\mar{b3205}\beq
\G=dx^\la\ot(\dr_\la + \frac12A_\la{}^{ab}
L_{ab}{}^c{}_dh^d_\m\dot x^\m h_c^\nu\dot\dr_\nu) \label{b3205}
\eeq
where $L_{ab}{}^c{}_d= \eta_{bd}\dl^c_a - \eta_{ad}\dl^c_b$ are
generators of a right Lie algebra $\ccG_\rL$ of a proper Lorentz
group L in a Minkowski space $\Bbb R^4$. Written relative to a
holonomic atlas, the connection $\G$ (\ref{b3205}) possesses
components
\mar{mos190}\beq
\G_\la{}^\m{}_\nu = h^k_\nu\dr_\la h^\m_k + \eta_{ka}h^\m_b
h^k_\nu A_\la{}^{ab}. \label{mos190}
\eeq

\begin{theo} Any Lorentz connections is a metric connection for
some pseudo-Riemannian metric $g$, and {\it vice versa}.
\end{theo}

\begin{proof}
By virtue of the well-known theorem \cite{book09,kob}, a metric
connection $\G$ for a pseudo-Riemmanian metric $g=\zeta\circ h$ is
reducible to a Lorentz connection on the proper Lorentz subbundle
$L^hX$, i.e., it is a Lorentz connection.  Conversely, every
Lorentz connection obeys the metricity condition (\ref{mos203})
for a pseudo-Riemannian metric $g=\zeta\circ h$.
\end{proof}

\begin{theo} \label{gtg} \mar{gtg}
Any world connection $\G$ defines a Lorentz connection $\G_h$ on
each principal L-subbundle $L^hX$ of a frame bundle.
\end{theo}

\begin{proof}
The Lie algebra of $GL_4$ is a direct sum
\mar{g13}\beq
\ccG_{GL_4} = \ccG_{\rL} \oplus {\got m} \label{g13}
\eeq
of the Lie algebra $\ccG_{\rL}$ of a Lorentz group and a subspace
${\got m}$ such that $[\ccG_{\rL},{\got m}]\subset {\got m}$.
Therefore, let us consider a local connection one-form of a
connection $\G$ with respect to a Lorentz atlas $\Psi^h$ of $LX$
given by a tetrad forms $h^a$. It reads
\be
z_\iota^{h*}\ol \G=- \G_\la{}^b{}_a dx^\la\ot L_b{}^a,\qquad
\G_\la{}^b{}_a = -h^b_\m \dr_\la h^\m_a  + \G_\la{}^\m{}_\nu
h^b_\m h^\nu_a,
\ee
where $\{L_b^a\}$ is a basis for a Lie algebra $\ccG_{GL_4}$. The
Lorentz part of this form is precisely a local connection one-form
of a connection $\G_h$ on $L^hX$. We have
\mar{K102}\beq
z_\zeta^{h*}\ol \G_h= -\frac12 A_\la{}^{ab}dx^\la\ot L_{ab},
\qquad A_\la{}^{ab} =\frac12
(\eta^{kb}h^a_\m-\eta^{ka}h^b_\m)(\dr_\la h^\m_k -
 h^\nu_k \G_\la{}^\m{}_\nu). \label{K102}
\eeq
Then combining this expression and the expression (\ref{b3205})
gives the connection
\mar{a3205}\beq
\G_h=dx^\la\ot(\dr_\la +
\frac14(\eta^{kb}h^a_\m-\eta^{ka}h^b_\m)(\dr_\la h^\m_k - h^\nu_k
\G_\la{}^\m{}_\nu) L_{ab}{}^c{}_dh^d_\m\dot x^\m
h_c^\nu\dot\dr_\nu) \label{a3205}
\eeq
with respect to a Lorentz atlas $\Psi^h$ and this connection
\mar{ppp}\beq
\G_h=dx^\la\ot[\dr_\la
+\frac12(h^k_\al\dl^\bt_\m-\eta^{kc}g_{\m\al}h^\bt_c) (\dr_\la
h^\m_k - h^\nu_k \G_\la{}^\m{}_\nu)\dot x^\al\dr_\bt] \label{ppp}
\eeq
relative to a holonomic atlas. If $\G$ is the Lorentz connection
(\ref{mos190}) extended from $L^hX$, then obviously $\G_h=\G$.
\end{proof}

\section{Space-time structure}

If a structure group $GL_4$ (\ref{gl4}) of a frame bundle $LX$ is
reducible to a proper Lorentz group L, it is always reducible to
the maximal compact subgroup $SO(3)$ of L. A structure group
$GL_4$ of $LX$ also is reducible to its maximal compact subgroup
$SO(4)$. Thus, there is the commutative diagram
\mar{8}\beq
\begin{array}{ccc}
 GL_4 &  \longrightarrow & SO(4)  \\
\put(0,10){\vector(0,-1){20}} & & \put(0,10){\vector(0,-1){20}} \\
 {\rm L} & \longrightarrow & SO(3)
\end{array} \label{8}
\eeq
of the reduction of structure groups of a frame bundle $LX$ in
gravitation theory \cite{sardz}. This reduction diagram results in
the following.

(i) There is one-to-one correspondence between the reduced
principal $SO(4)$-subbundles $L^{g^R}X$ of a frame bundle $LX$ and
the global sections of the quotient bundle $LX/SO(4)\to X$. Its
global sections are Riemannian metrics $g^R$ on $X$. Thus, a
Riemannian metric on a world manifold always exists.

(ii) As was mentioned above, a reduction of a structure group of a
frame bundle $LX$ to a proper Lorentz group means the existence of
a reduced proper Lorentz subbundle $L^hX\subset LX$ associated
with a tetrad field $h$ or a pseudo-Riemannian metric
$g=\zeta\circ h$ on $X$.

(iii) Since a structure group $L$ of this reduced Lorentz bundle
$L^hX$ is reducible to a group $SO(3)$, there exists a reduced
principal $SO(3)$-subbundle
\mar{spat}\beq
L^h_0X\subset L^hX\subset LX, \label{spat}
\eeq
called the spatial structure. The corresponding global section of
a quotient bundle $L^hX/SO(3)\to X$ with a typical fibre $\Bbb
R^3$ is a one-codimensional spatial distribution $\bF\subset TX$
on $X$. Its annihilator is a one-dimensional codistribution
$\bF^*\subset T^*X$.

Given the spatial structure $L^h_0X$ (\ref{spat}), let us consider
the Lorentz bundle atlas $\Psi^h_0$ (\ref{lat}) given by local
sections $z_\iota$ of $LX$ taking their values into a reduced
$SO(3)$-subbundle $L^h_0X$. Its transition functions are
$SO(3)$-valued.

It follows that, in gravitation theory on a world manifold $X$,
one can always choose an atlas of the tangent bundle $TX$ and
associated bundles with $SO(3)$-valued transition functions.  This
bundle atlas, called the spatial bundle atlas.

Given a spatial bundle atlas $\Psi^h_0$, its $SO(3)$-valued
transition functions preserve a time-like component
\mar{h0a}\beq
h^0=h^0_\la dx^\la \label{h0a}
\eeq
of local tetrad forms (\ref{b3211}) which, therefore, is globally
defined. We agree to call it the time-like tetrad form.
Accordingly, the dual time-like vector field
\mar{h0b}\beq
h_0=h^\m_0\dr_\m \label{h0b}
\eeq
also is globally defined. In this case, a spatial distribution
$\bF$ is spanned by spatial components $h_i$, $i=1,2,3$, of the
tetrad frames (\ref{b3211a}), while the time-like tetrad form
(\ref{h0a}) spans the tetrad codistribution $\bF^*$, i.e.,
\mar{yuy2}\beq
h^0\rfloor \bF=0. \label{yuy2}
\eeq
Then the tangent bundle $TX$ of a world manifold $X$ admits a
space-time decomposition
\mar{mos209}\beq
TX=\bF\oplus T^0X, \label{mos209}
\eeq
where $T^0X$ is a one-dimensional fibre bundle spanned by the
time-like vector field $h_0$ (\ref{h0b}).

Due to the commutative diagram (\ref{8}), the reduced L-subbundle
$L^h_0X$ (\ref{spat}) of a reduced Lorentz bundle $L^hX$ is a
reduced subbundle of some reduced $SO(4)$-bundle $L^{g^R}X$ too,
i.e.,
\mar{yuy}\beq
L^hX\supset L^h_0X\subset L^{g^R}X. \label{yuy}
\eeq
Let $g=\zeta\circ h$ and $g^R$ be the corresponding
pseudo-Riemannian and Riemannian metrics on $X$. Written with
respect to a spatial bundle atlas $\Psi^h_0$, they read
\mar{prm1,2}\ben
&& g=\eta_{ab}h^a\ot h^b, \qquad
g_{\m\nu}=h_\m^ah_\nu^b\eta^{ab}, \label{prm1}\\
&& g^R=\eta^E_{ab}h^a\ot h^b, \qquad
g^R_{\m\nu}=h_\m^ah_\nu^b\eta^E_{ab}, \label{prm2}
\een
where $\eta^E$ is an Euclidean metric in $\Bbb R^4$. The
space-time decomposition (\ref{mos209}) is orthonormal with
respect to both the metrics (\ref{prm1}) and (\ref{prm2}). The
following well-known theorem holds \cite{book09,haw}.

\begin{theo} \label{sptime} \mar{sptime}
For any pseudo-Riemannian metric $g$ on a world manifold $X$,
there exist a normalized time-like one-form $h^0$ and a Riemannian
metric $g^R$ such that
\mar{mos208}\beq
g=2h^0\ot h^0 -g^R. \label{mos208}
\eeq
Conversely, let a world manifold $X$ admit a nowhere vanishing
one-form $\si$ (or, equivalently, a nowhere vanishing vector
field). Then  any Riemannian world metric $g^R$ on $X$ yields the
pseudo-Riemannian world metric $g$ (\ref{mos208}) where
$h^0=\si(g^R(\si,\si))^{-1/2}$.
\end{theo}

\begin{cor} A world manifold $X$ admits a pseudo-Riemannian
metric iff there exists a nowhere vanishing one-form (or a vector
field) on $X$.
\end{cor}

Note that the condition (\ref{yuy}) gives something more.

\begin{theo} \label{yuy1} \mar{yuy1}
There  is  one-to-one correspondence between the reduced
$SO(3)$-subbundles of a frame bundle $LX$ and the triples
$(g,\bF,g^R)$  of  a pseudo-Riemannian metric $g$, a spatial
distribution $\bF$ defined by the condition (\ref{yuy2}) and a
Riemannian metric $g^R$ which obey the relation (\ref{mos208}).
\end{theo}

A spatial distribution $\bF$ and a Riemannian metric $g^R$ in the
triple $(g,\bF,g^R)$ in Theorem \ref{yuy1} are called
$g$-compatible. The corresponding space-time decomposition is said
to be a $g$-compatible space-time structure. A world manifold
endowed with a pseudo-Riemannian metric and a compatible
space-time structure is called the space-time.

\begin{rem}
A $g$-compatible Riemannian metric  $g^R$   in  a triple
$(g,\bF,g^R)$ defines a $g$-compatible distance function $d(x,x')$
on a world manifold $X$.  Such a function brings $X$ into a metric
space whose locally Euclidean topology is equivalent to a manifold
topology on $X$. Given a gravitational field $g$, the
$g$-compatible Riemannian metrics and the corresponding distance
functions are different for different spatial distributions $\bF$
and $\bF'$.  It  follows that physical observers associated  with
different spatial distributions $\bF$ and $\bF'$ perceive a world
manifold $X$ as different Riemannian spaces. The well-known
relativistic changes of sizes of moving bodies exemplify this
phenomenon \cite{sardz}.
\end{rem}

A space-time structure is called integrable if a spatial
distribution $\bF$, given by the condition (\ref{yuy2}), is
involutive. In this case, its integral manifolds constitute a
spatial foliation $\cF$ of a world manifold $X$ whose leaves are
spatial three-dimensional subspaces of $X$. A spatial distribution
$\bF$ is integrable iff the one-form $h^0$ (\ref{h0a}) obeys the
condition $dh^0\w h^0=0$. In this case, the time-like vector field
$h_0$ (\ref{h0b}) is transversal to a spatial foliation $\cF$.

A spatial foliation $\cF$ is called causal if no curve transversal
to its leaves intersects each leave more than once. This condition
is equivalent to the stable causality of Hawking \cite{haw}.

\begin{theo} \label{t2} \mar{t2} A space-time foliation $\cF$
is causal iff it is a foliation of level surfaces of some smooth
real function $f$ on $X$ whose differential nowhere vanishes. Such
a foliation is simple, i.e., there is a fibred manifold $f:X\to
\Bbb R$ whose fibres are leaves of $\cF$.
\end{theo}

This is not the case of a compact world manifold which can not be
a fibred manifold over $\Bbb R$. Thus, a compact world manifold
fails to satisfy the stable causality condition.

The stable causality does not provide the simplest causal
structure. If a fibred manifold $X\to\Bbb R$ in Theorem \ref{t2}
is a fibre bundle, it is trivial, i.e., a world manifold $X$ is a
globally hyperbolic space $X=\Bbb R \times M$. Since any oriented
three-dimensional manifold is parallelizable, a globally
hyperbolic space also is parallelizable.

\section{Metric-affine gravitation theory}

In the absence of matter fields, dynamic variables of
metric-affine gravitation theory are world connections and
pseudo-Riemannian metrics on $X$. Their Lagrangian $L_{\rm MA}$ is
invariant under general covariant transformations.

\begin{rem} In view of the decomposition (\ref{mos191}), one can
choose a different collection of dynamic variables of
metric-affine gravitation theory. These are a pseudo-Riemannian
metric, a torsion and the non-metricity tensor (\ref{mos193}). A
problem is that, in this case, we deal with differential equations
of fourth and higher order in a pseudo-Riemannian metric, unless
$L_{\rm MA}$ is the Hilbert--Einstein Lagrangian $L_{\rm HE}$
(\ref{10221}).
\end{rem}

World connections are represented by sections of the bundle of
world connections $C_{\rm W}$ (\ref{015}). Pseudo-Riemannian world
metrics are described by sections of the quotient bundle
(\ref{b3203}). Therefore, let us consider the bundle product
\mar{ggtt}\beq
Y=\Si_{\rm PR}\op\times_X C_{\rm W} \label{ggtt}
\eeq
coordinated by $(x^\la,\si^{\m\nu}, k_\mu{}^\al{}_\bt)$.

Let us restrict our consideration to first order Lagrangian theory
on $Y$ (\ref{ggtt}). In this case, a configuration space of gauge
gravitation theory is the jet manifold
\mar{kkl}\beq
J^1Y= J^1\Si_{\rm PR}\op\times_X J^1C_{\rm W}, \label{kkl}
\eeq
coordinated by $(x^\la,\si^{\m\nu}, k_\mu{}^\al{}_\bt,
\si_\la^{\m\nu}, k_{\la\mu}{}^\al{}_\bt)$.

A first order Lagrangian $L_{\rm MA}$ of metric-affine gravitation
theory is a defined as a density
\mar{10130}\beq
L_{\rm AM}=\cL_{\rm AM}(x^\la,\si^{\m\nu}, k_\mu{}^\al{}_\bt,
\si_\la^{\m\nu}, k_{\la\mu}{}^\al{}_\bt)\om, \qquad
\om=dx^1\w\cdots\w dx^4, \label{10130}
\eeq
on the configuration space $J^1Y$ (\ref{kkl}) \cite{book09}. Its
Euler--Lagrange operator takes the form
\mar{eeu}\ben
&& \dl L_{\rm MA}= (\cE_{\al\bt} d\si^{\al\bt} + \cE^\m{}_\al{}^\bt
dk_\m{}^\al{}_\bt)\w \om. \label{eeu} \\
&& \cE_{\al\bt} =\left(\frac{\dr}{\dr \si^{\al\bt}}
 - d_\la\frac{\dr}{\dr \si^{\al\bt}_\la}\right)\cL_{\rm AM},\qquad
\cE^\m{}_\al{}^\bt =\left(\frac{\dr}{\dr k_\mu{}^\al{}_\bt}
 - d_\la\frac{\dr}{\dr k_{\la\mu}{}^\al{}_\bt}\right)\cL_{\rm AM},
 \nonumber\\
&& d_\la=\dr_\la + \si^{\al\bt}_\la \frac{\dr}{\dr \si^{\al\bt}} +
k_{\la\mu}{}^\al{}_\bt \frac{\dr}{\dr k_\mu{}^\al{}_\bt} +
\si^{\al\bt}_{\la\nu} \frac{\dr}{\dr \si^{\al\bt}_\nu} +
k_{\la\nu\mu}{}^\al{}_\bt \frac{\dr}{\dr k_{\nu\mu}{}^\al{}_\bt}.
\nonumber
\een
The corresponding Euler--Lagrange equations read
\mar{10141}\beq
\cE_{\al\bt}=0, \qquad \cE^\m{}_\al{}^\bt=0. \label{10141}
\eeq

\begin{rem} \label{hhtuu} \mar{hhtuu}
The  Hilbert--Einstein Lagrangian $L_{\rm HE}$ (\ref{10201}) of
General Relativity depends only on metric variables
$\si^{\al\bt}$. It is a reduced second order Lagrangian which
differs from the first order one $L'_{\rm HE}$ in a variationally
trivial term and leads to second order Euler--Lagrange equations
(\ref{10205}).
\end{rem}

The fibre bundle (\ref{ggtt}) is a natural bundle admitting the
functorial lift
\mar{gr3}\ben
&& \wt\tau_{\Si C}=\tau^\m\dr_\m +(\si^{\nu\bt}\dr_\nu \tau^\al
+\si^{\al\nu}\dr_\nu \tau^\bt)\frac{\dr}{\dr \si^{\al\bt}} +
\label{gr3}\\
&& \qquad (\dr_\nu \tau^\al k_\m{}^\nu{}_\bt -\dr_\bt \tau^\nu
k_\m{}^\al{}_\nu -\dr_\mu \tau^\nu k_\nu{}^\al{}_\bt
+\dr_{\m\bt}\tau^\al)\frac{\dr}{\dr k_\mu{}^\al{}_\bt} \nonumber
\een
of vector fields $\tau$ on $X$ \cite{book09}. It is an
infinitesimal generator of general covariant transformations. At
the same time, $\wt\tau_{\Si C}$ (\ref{gr3}) also is an
infinitesimal gauge transformation whose gauge parameters are
components $\tau^\la(x)$ of vector fields $\tau$ on $X$.

By virtue of the relativity principle, the Lagrangian $L_{\rm MA}$
(\ref{10130}) of metric-affine gravitation theory is assumed to be
invariant under general covariant transformations. Its Lie
derivative along the jet prolongation $J^1\wt\tau_{\Si C}$ of the
vector field $\wt\tau_{\Si C}$ for any $\tau$ vanishes, i.e.,
\mar{10140}\beq
\bL_{J^1\wt\tau_{\Si C}}L_{\rm AM}=0. \label{10140}
\eeq

\begin{rem}
The total group of automorphisms of a frame bundle $LX$ also is
considered in gauge gravitation theory \cite{heh}. As was
mentioned above, such an automorphism is a composition of some
general covariant transformation and a vertical automorphism of
$LX$.  A problem is that the most of gravitation Lagrangians,
e.g., the Hilbert--Einstein Lagrangian are not invariant under
vertical frame transformations. To overcome this difficulty, one
considers frame fields (i.e., sections of a frame bundle $LX$) as
dynamic variables. However, these fields fail to be global, unless
a world manifold $X$ is parallelizable.
\end{rem}

As was mentioned above a configuration space $J^1C_{\rm W}$ of
world connections possesses the canonical splitting (\ref{0101}).
The following assertion is analogous to the well-known Utiyama
theorem in Yang--Mills gauge theory of principal connections
\cite{book09,sard08}.

\begin{theo} \label{httu1} \mar{httu1} If a first order Lagrangian $L_{\rm MA}$
on the configuration space (\ref{kkl}) is invariant under general
covariant transformations and it does not depend on the jet
coordinates $\si^{\al\bt}_\la$ (i.e., derivatives of a metric),
this Lagrangian factorizes through the terms
$\cR_{\la\m}{}^\al{}_\bt$ (\ref{0101}).
\end{theo}

In contrast with the well-known Lagrangian of Yang--Mills gauge
theory, different contractions of a curvature tensor
$\cR_{\la\m}{}^\al{}_\bt$ are possible. For instance, the Ricci
tensor $R_c$ (\ref{ric}) and a scalar curvature
$\cR=\si^{\m\nu}\cR_{\la\m}{}^\la{}_\nu$ are defined. Moreover, a
Lagrangian $L_{\rm MA}$ also can depend on a torsion
\mar{10145}\beq
t_\m{}^\nu{}_\la = k_\m{}^\nu{}_\la - k_\la{}^\nu{}_\m.
\label{10145}
\eeq
The Yang--Mills Lagrangian in metric-affine gravitation theory is
given by the expression
\be
\si^{\m\la}\si^{\nu\g}\cR_{\m\nu}{}^\al{}_\bt
\cR_{\la\g}{}^\bt{}_\al.
\ee
It is invariant under the total group of automorphisms of a frame
bundle $LX$. In this case, metric variables $\si^{\m\la}$ fail to
be dynamic because they are brought into a constant Minkowski
metric by general frame transformations.

\section{Energy-momentum conservation law}

Since infinitesimal general covariant transformations
$\wt\tau_{\Si C}$ (\ref{gr3}) are exact symmetries of a
metric-affine gravitation Lagrangian, let us study the
corresponding conservation law. This is the energy-momentum
conservation laws because vector fields $\wt\tau_{\Si C}$ are not
vertical \cite{book09,sard97}.  Moreover, since infinitesimal
general covariant transformations $\wt\tau_{\Si C}$ (\ref{gr3})
are infinitesimal gauge transformations depending on derivatives
of gauge parameters, the corresponding energy-momentum current
reduces to a superpotential \cite{book09,sard09}.

In view of Theorem \ref{httu1}, let us assume that a metric-affine
gravitation Lagrangian $L_{\rm MA}$ is independent of the
derivative coordinates $\si_\la{}^{\al\bt}$ of a world metric and
that it factorizes through the curvature terms
$\cR_{\la\m}{}^\al{}_\bt$ (\ref{0101}). Then the following
relations take place:
\mar{K300',}\ben
&&  \pi^{\la\nu}{}_\al{}^\bt= -\pi^{\nu\la}{}_\al{}^\bt, \qquad
\pi^{\la\nu}{}_\al{}^\bt=\frac{\dr \cL_{\rm MA}}{\dr k_{\la\nu}{}^\al{}_\bt}, \label{K300'}\\
&&\frac{\dr\cL_{\rm MA}}{\dr k_\nu{}^\al{}_\bt}=
\pi^{\la\nu}{}_\al{}^\si k_\la{}^\bt{}_\si
-\pi^{\la\nu}{}_\si{}^\bt k_\la{}^\si{}_\al. \label{K300}
\een

Let us follow the compact notation
\be
y^A=k_\m{}^\al{}_\bt, \qquad u_\m{}^\al{}_\bt{}^{\ve\si}_\g =
\dl^\ve_\m \dl^\si_\bt \dl^\al_\g, \qquad
u_\m{}^\al{}_\bt{}^\ve_\g= k_\m{}^\ve{}_\bt \dl^\al_\g
-k_\m{}^\al{}_\g \dl^\ve_\bt - k_\g{}^\al{}_\bt \dl^\ve_\m.
\ee
Then the vector field (\ref{gr3}) takes the form
\be
\wt\tau_{\Si C} =\tau^\la\dr_\la  + (\si^{\nu\bt}\dr_\nu\tau^\al
+\si^{\al\nu}\dr_\nu\tau^\bt)\dr_{\al\bt}+
(u^A{}_\al^\bt\dr_\bt\tau^\al
+u^A{}_\al^{\bt\m}\dr_{\bt\m}\tau^\al)\dr_A.
\ee
We also have the equalities
\be
\pi^\la_A u^A{}_\al^{\bt\m} =\pi^{\la\m}{}_\al{}^\bt,\qquad
\pi^\ve_A u^A{}_\al^\bt = -\dr^\ve{}_\al{}^\bt\cL_{\rm MA} -
\pi^{\ve\bt}{}_\si{}^\g k_\al{}^\si{}_\g.
\ee

Let $L_{\rm MA}$ be invariant under general covariant
transformations, i.e., the equality (\ref{10140}) for any vector
field $\tau$ holds. Then the first variational formula leads to an
equality
\mar{J4a}\ben
&& 0= (\si^{\nu\bt}\dr_\nu\tau^\al +\si^{\al\nu}\dr_\nu\tau^\bt -\tau^\la\si^{\al\bt}_\la)
\dl_{\al\bt}\cL_{\rm MA} +
\label{J4a}\\
&& \qquad (u^A{}_\al^\bt\dr_\bt\tau^\al
+u^A{}_\al^{\bt\m}\dr_{\bt\m}\tau^\al - \tau^\la y^A_\la) \dl_A
\cL_{\rm MA} -\nonumber\\
&& \qquad d_\la[ \pi^\la_A(y^A_\al\tau^\al -u^A{}_\al^\bt\dr_\bt\tau^\al
 -u^A{}_\al^{\ve\bt}\dr_{\ve\bt}\tau^\al) -\tau^\la\cL_{\rm MA}].
 \nonumber
\een
This equality on the shell (\ref{10141}) results in the weak
conservation law
\mar{K8}\ben
&& 0\ap - d_\la[ \pi^\la_A(y^A_\al\tau^\al -u^A{}_\al^\bt\dr_\bt\tau^\al
 -u^A{}_\al^{\ve\bt}\dr_{\ve\bt}\tau^\al) -\tau^\la\cL_{\rm MA}],\label{K8}
\een
of the energy-momentum current of metric-affine gravity
\mar{b3190}\beq
\cJ_{\rm MA}{}^\la= \pi^\la_A(y^A_\al\tau^\al
-u^A{}_\al^\bt\dr_\bt\tau^\al
 -u^A{}_\al^{\ve\bt}\dr_{\ve\bt}\tau^\al)-\tau^\la\cL_{\rm MA}. \label{b3190}
\eeq

\begin{rem}
It is readily observed that, with respect to a local coordinate
system where a vector field $\tau$ is constant, the
energy-momentum current (\ref{b3190}) leads to the canonical
energy-momentum tensor
\be
\cJ_{\rm MA}{}^\la{}_\al\tau^\al =(\pi^{\la\m}{}_\bt{}^\nu
k_{\al\m}{}^\bt{}_\nu -\dl^\la_\al \cL_{\rm MA})\tau^\al,
\ee
suggested in order to describe an energy-momentum complex in the
Palatini model \cite{dick}.
\end{rem}

Due to the arbitrariness of $\tau^\la$, the equality (\ref{J4a})
falls into a set of equalities
\mar{b3173d,-b}\ben
&& \pi^{(\la\ve}{}_\g{}^{\si)}=0, \label{b3173d}\\
&& (u^A{}_\g^{\ve\si}\dr_A + u^A{}_\g^\ve\dr^\si_A)\cL_{\rm MA}= 0, \label{b3173c}\\
&& \dl^\bt_\al\cL_{\rm MA} + 2\si^{\bt\m}\dl_{\al\m}\cL_{\rm MA} + u^A{}_\al^\bt\dl_A\cL_{\rm MA}
 + d_\m(\pi^\m_A  u^A{}_\al^\bt)
-y^A_\al\pi^\bt_A  =0,\label{b3173b} \\
&& \dr_\la\cL_{\rm MA}=0. \nonumber
\een

\begin{rem} It is readily observed that
the equalities (\ref{b3173d}) and (\ref{b3173c}) hold due to the
relations (\ref{K300'}) and (\ref{K300}), respectively.
\end{rem}

Substituting the term $y^A_\al\pi^\bt_A$ from the expression
(\ref{b3173b}) in the energy-momentum conservation law (\ref{K8}),
one brings this conservation law into the form
\mar{b3174}\ben
&& 0\ap -
d_\la[2\si^{\la\m}\tau^\al\dl_{\al\m}\cL_{\rm MA} +
u^A{}_\al^\la\tau^\al\dl_A\cL_{\rm MA} - \pi^\la_Au^A{}_\al^\bt\dr_\bt\tau^\al + \label{b3174}\\
&& \qquad d_\m(\pi^{\la\m}{}_\al{}^\bt)
\dr_\bt\tau^\al + d_\m(\pi^\m_A  u^A{}_\al^\la)\tau^\al -
d_\m(\pi^{\la\m}{}_\al{}^\bt \dr_\bt\tau^\al)]. \nonumber
\een
After separating the variational derivatives, the energy-momentum
conservation law (\ref{b3174}) of a metric-affine gravity takes a
superpotential form
\be
&& 0\ap - d_\la [2\si^{\la\m}\tau^\al\dl_{\al\m}\cL_{\rm MA}
+(k_\m{}^\la{}_\g\dl^\m{}_\al{}^\g\cL_{\rm MA} -
 k_\m{}^\si{}_\al\dl^\m{}_\si{}^\la\cL_{\rm MA} -
k_\al{}^\si{}_\g\dl^\la{}_\si{}^\g\cL_{\rm MA})\tau^\al +  \\
&& \qquad \dl^\la{}_\al{}^\m\cL_{\rm MA}\dr_\m\tau^\al
-d_\m(\dl^\m{}_\al{}^\la\cL_{\rm MA})\tau^\al +
 d_\m(\pi^{\m\la}{}_\al{}^\nu(\dr_\nu\tau^\al
-k_\si{}^\al{}_\nu\tau^\si))],
\ee
where an energy-momentum current on-shell reduces to a generalized
Komar superpotential
\mar{K3}\beq
 U_{\rm MA}{}^{\m\la}= 2\frac{\dr\cL_{\rm MA}}{\dr
\cR_{\m\la}{}^\al{}_\nu}(D_\nu\tau^\al +
t_\nu{}^\al{}_\si\tau^\si), \label{K3}
\eeq
where $D_\nu$ is a covariant derivative relative to a connection
$k_\nu{}^\al{}_\si$ and $t_\nu{}^\al{}_\si$ is its torsion
\cite{giacqg,sard97b}.

\section{General Relativity}

In Einstein's General Relativity, dynamic variables are only a
pseudo-Riemannian metric $\si^{\m\nu}$, while a world connection
is restricted to the Levi--Civita one
\mar{07103}\beq
k_\m{}^\bt{}_\la = \{_\m{}^\bt{}_\la\}=
-\frac12\si^{\bt\nu}(d_\m\si_{\nu\la} + d_\la\si_{\m\nu} -
d_\nu\si_{\m\la}). \label{07103}
\eeq

A Lagrangian of General Relativity is the Hilbert--Einstein
Lagrangian
\mar{10201}\ben
&& L_{\rm HE}=\frac12\cR\sqrt{\si}\om= \si^{\m\bt} \cR_{\la\m}{}^\la{}_{\bt}\sqrt{\si}\om, \qquad
\si=|{\rm \,det}(\si_{\al\bt})|, \label{10201}\\
&& \cR_{\la\m}{}^\al{}_\bt = d_\la\{_\m{}^\al{}_\bt\} - d_\m
\{_\la{}^\al{}_\bt\} + \{_\la{}^\g{}_\bt\} \{_\m{}^\al{}_\g\} -
\{_\m{}^\g{}_\bt\} \{_\la{}^\al{}_\g\}. \nonumber
\een
It is a reduced second order Lagrangian resulting in second order
Euler--Lagrange equations
\mar{10205}\beq
\cE_{\al\bt}=\cR_{\al\bt} -\frac12 \si_{\al\bt} \cR=0.
\label{10205}
\eeq
The Hilbert--Einstein Lagrangian differs from the first order one
$L'_{\rm HE}$ in a variationally trivial term $d_\la f^\la\om$.
General covariant transformations are variational, but not exact
symmetries of a Lagrangian $L'_{\rm HE}$.

In metric-affine gravitation theory, the Hilbert--Einstein
Lagrangian takes the form
\mar{10221}\ben
&& L=\cR\sqrt{\si}\om =\si^{\m\bt} \cR_{\la\m}{}^\la{}_{\bt}\sqrt{\si}\om, \label{10221}\\
&& \cR_{\la\m}{}^\al{}_\bt = d_\la k_\m{}^\al{}_\bt - d_\m
k_\la{}^\al{}_\bt + k_\la{}^\g{}_\bt k_\m{}^\al{}_\g -
k_\m{}^\g{}_\bt k_\la{}^\al{}_\g, \nonumber
\een
where $\cR$ is a scalar curvature. The corresponding
Euler--Lagrange equations read
\mar{10180,1}\ben
&& \cE_{\al\bt}=\cR_{\al\bt} -\frac12 \si_{\al\bt} \cR=0, \label{10180}\\
&& \cE^\nu{}_\al{}^\bt=-d_\al(\si^{\nu\bt} \sqrt{\si})
+d_\la(\si^{\la\bt}\sqrt{\si})\dl^\nu_\al + \label{10181}\\
&& \qquad (\si^{\nu\g} k_\al{}^\bt{}_\g -\si^{\la\g}\dl^\nu_\al
k_\la{}^\bt{}_\g - \si^{\nu\bt}k_\g{}^\g{}_\al
+\si^{\la\bt}k_\la{}^\nu{}_\al)\sqrt{\si}=0. \nonumber
\een
The equation (\ref{10180}) is an analogy of the Einstein
equations, whereas the equations (\ref{10181}) describe the
torsion $t_\m{}^\nu{}_\la$ (\ref{10145}) and the non-metricity
\be
c_{\m\nu\al}=c_{\m\al\nu}=d_\m \si_{\nu\al}
+k_\m{}^\bt{}_\al\si_{\nu\bt} + k_\m{}^\bt{}_\nu\si_{\bt\al}
\ee
of a world connection. They are brought into the form
\mar{10209}\ben
&& \sqrt{\si^{-1}}\si_{\nu\ve}\si_{\bt\m}\cE^\nu{}_\al{}^\bt=
 c_{\al\ve\m}-\frac12 \si_{\m\ve}\si^{\la\g}c_{\al\la\g} -
\si_{\al\ve}\si^{\la\bt}c_{\la\bt\m} + \label{10209} \\
&& \qquad \frac12
\si_{\al\ve}\si^{\la\g}c_{\m\la\g} +
 t_{\m\ve\al} + \si_{\m\ve} t_\al{}^\g{}_\g + \si_{\al\ve} t_\g{}^\g{}_\m
=0. \nonumber
\een

The Hilbert--Einstein Lagrangian (\ref{10221}) is invariant under
general covariant transformations. The corresponding generalized
Komar superpotential (\ref{K3}) comes to the well-known Komar
superpotential if one substitutes the Levi--Civita connection
$k_\nu{}^\al{}_\si =\{_\nu{}^\al{}_\si\}$ (\ref{07103}).

\section{BRST extension}

A BRST extension of Lagrangian field theory is a first step
towards its quantization \cite{book09,sard08}.

Taking the vertical part of vector fields $\wt\tau_{\Si C}$
(\ref{gr3}) and replacing gauge parameters $\tau^\la$ with ghosts
$c^\la$, we obtain the odd vertical graded derivation
\mar{nbnm}\ben
&&u=u^{\al\bt}\frac{\dr}{\dr\si^{\al\bt}} +u_\m{}^\al{}_\bt
\frac{\dr}{\dr k_\mu{}^\al{}_\bt} = (\si^{\nu\bt} c_\nu^\al
+\si^{\al\nu} c_\nu^\bt-c^\la\si_\la^{\al\bt})\frac{\dr}{\dr
\si^{\al\bt}}+
\label{nbnm}\\
&& \qquad (c_\nu^\al k_\m{}^\nu{}_\bt -c_\bt^\nu k_\m{}^\al{}_\nu
-c_\mu^\nu k_\nu{}^\al{}_\bt +c_{\m\bt}^\al-c^\la
k_{\la\mu}{}^\al{}_\bt)\frac{\dr}{\dr k_\mu{}^\al{}_\bt}.
\nonumber
\een

Since the infinitesimal gauge transformations $\wt\tau_{\Si C}$
(\ref{gr3}) are exact symmetries of a metric-affine Lagrangian
$L_{\rm MA}$, the vertical graded derivation $u$ (\ref{nbnm}) is a
variational symmetry of $L_{\rm MA}$ and, thus, is its gauge
symmetry. As a consequence, the Euler--Lagrange operator $\dl
L_{\rm MA}$ (\ref{eeu}) of this Lagrangian obeys the complete
Noether identities
\mar{333}\ben
&& -\si^{\al\bt}_\la \cE_{\al\bt} - 2d_\m(\si^{\m\bt}\cE_{\la\bt} -
k_{\la\m}{}^\al{}_\bt\cE^\m{}_\al{}^\bt- \label{333}\\
&&\qquad d_\m[(k_\nu{}^\m{}_\bt\dl^\al_\la - k_\nu{}^\al{}_\la \dl^\m_\bt
- k_\la{}^\al{}_\bt \dl^\m_\nu)\cE^\nu{}_\al{}^\bt] +
d_{\m\bt}\cE^\m{}_\la{}^\bt=0.\nonumber
\een
These Noether identities are irreducible. Therefore, the graded
derivation (\ref{nbnm}) is the gauge operator of gauge gravitation
theory. It admits the nilpotent BRST extension
\mar{444}\beq
{\bf c}=u + c^\la_\m c^\m\frac{\dr}{\dr c^\la} \label{444}
\eeq
(which differs from that in \cite{gron}).

Accordingly, an original gravitation Lagrangian $L_{\rm MA}$
admits a BRST extension to a proper solution of the master
equation which reads
\mar{555}\beq
L_E= L_{\rm MA} + u^{\al\bt}\ol\si_{\al\bt}\om + u_\m{}^\al{}_\bt
\ol k^\m{}_\al{}^\bt\om + c^\la_\m c^\m \ol c_\la\om, \label{555}
\eeq
where $\ol\si_{\al\bt}$, $\ol k^\m{}_\al{}^\bt$ and $\ol c_\la$
are the corresponding antifields \cite{gauge05,book09}.

\section{Dirac spinor fields}

Apparently, the existence of Dirac's fermion matter possessing
Lorentz symmetries is an underlying physical reason of an
appearance of a Lorentz reduced structure and, consequently, a
(tetrad) gravitational field.

Dirac spinors are conventionally described in the framework of
formalism of Clifford algebras \cite{law}.

Let $M=\Bbb R^4$ be a Minkowski space equipped with a Minkowski
metric $\eta$. Let $\Bbb C_{1,3}$ be a complex Clifford algebra
generated by elements of $M$. It is a complexified quotient of the
tensor algebra of $M$ by a two-sided ideal spanned by elements
\be
e\otimes e'+e'\otimes e-2\eta(e,e')\in \ot M,\qquad e,e'\in M.
\ee

A Dirac spinor space $V$ is defined as a minimal left ideal of
$\Bbb C_{1,3}$ in which this algebra acts on the left. There is a
representation
\mar{w01}\beq
\g: M\otimes V \to V, \qquad \g(e^\al)=\g^\al, \label{w01}
\eeq
of elements of a Minkowski subspace $M\subset{\Bbb C}_{1,3}$ by
Dirac $\g$-matrices in $V$.

A Clifford group $G_{1,3}\subset \Bbb R_{1,3}$ is defined to
consist of invertible elements $l_s$ of a real Clifford algebra
$\Bbb R_{1,3}$ such that inner automorphisms given by these
elements preserve a Minkowski space $M\subset \Bbb R_{1,3}$, i.e.,
\beq
l_sel^{-1}_s = l(e), \qquad e\in M, \label{b3200}
\eeq
where $l$ is a Lorentz transformation of $M$. Hence, there is an
epimorphism of a Clifford group $G_{1,3}$ onto a Lorentz group
$O(1,3)$. However, the action (\ref{b3200}) of a Clifford group in
a Minkowski space $M$ is not effective.

A subgroup Pin$(1,3)$ of $G_{1,3}$ is generated by elements $e\in
M$ such that $\eta(e,e)=\pm 1$. An even part of Pin$(1,3)$ is a
spin group ${\rm Spin}(1,3)$, i.e., $\eta(e,e)=1$, $e\in {\rm
Spin}(1,3)$. Its component of the unity
\be
\rL_{\rm s}={\rm Spin}^0(1,3)\simeq SL(2,\Bbb C)
\ee
is the well-known two-fold universal covering group
\mar{b3204}\beq
z_L:\rL_{\rm s}\to \rL=\rL_{\rm s}/\Bbb Z_2  \label{b3204}
\eeq
of a proper Lorentz group L. We agree to call $\rL_{\rm s}$
(\ref{b3204}) the spinor Lorentz group. Its Lie algebra is that of
a proper Lorentz group L. We further consider an action of a
spinor Lorentz group $\rL_{\rm s}$, factorizing through that of a
proper Lorentz group L, in the Minkowski space $M$, but it is not
effective.

A Clifford group $G_{1,3}$ acts in a Dirac spinor space $V$ by
left multiplications $G_{1,3}\ni l_s:v\mapsto l_sv$, $v\in V$.
This action preserves the representation (\ref{w01}). A spinor
Lorentz group $\rL_{\rm s}$ acts in the Dirac spinor space $V$ by
means of the infinitesimal generators
\mar{b3213}\beq
L_{ab}=\frac{1}{4}[\g_a,\g_b]. \label{b3213}
\eeq

In classical field theory, Dirac spinor fields are described by
sections of a spinor bundle on a world manifold $X$ whose typical
fibre is the Dirac spinor space $V$ and whose structure group is
the spinor Lorentz group $\rL_{\rm s}$. In order to construct the
Dirac operator, one need a fibrewise action (\ref{w01}) of the
whole Clifford algebra $\Bbb C_{1,3}$ in a spinor bundle.
Therefore, a spinor bundle must be represented as a subbundle of
the bundle in Clifford algebras.

Let us start with a fibre bundle in Minkowski spaces $MX\to X$
over a world manifold $X$. It is defined as a fibre bundle with a
typical fibre $M$ and a structure group L. This fibre bundle is
extended to a fibre bundle in Clifford algebras $CX$ whose fibres
$C_xX$ are Clifford algebras generated by fibres $M_xX$ of a fibre
bundle in Minkowski spaces $MX$. A fibre bundle $CX$ possesses a
structure group of inner automorphisms of a complex Clifford
algebra ${\Bbb C}_{1,3}$. This structure group is reducible to a
proper Lorentz group L and, consequently, a bundle in Clifford
algebras $CX$ contains a subbundle $MX$ of the generating
Minkowski spaces. However, $CX$ need not contain a spinor
subbundle because a Dirac spinor subspace $V$ of ${\Bbb C}_{1,3}$
is not stable under inner automorphisms of ${\Bbb C}_{1,3}$. A
spinor subbundle $S_M$ of $CX$ exists if transition functions of
$CX$ can be lifted from a Clifford group $G_{1,3}$. This condition
agrees with the familiar condition of the existence of a spinor
structure.

A bundle $MX$ in Minkowski spaces must be isomorphic to the
cotangent bundle $T^*X$ in order that sections of a spinor bundle
$S_M$ describe spinor fields on a world manifold $X$.

A Dirac spinor structure on a world manifold $X$ is defined as a
pair $(P^h, z_s)$ of a principal bundle $P^h\to X$ with astructure
spin group $L_s=SL(2,\Bbb C)$ and its bundle morphism $z_s: P^h
\to LX$ to a frame bundle $LX$ \cite{law}. Any such morphism
factorizes
\mar{g10}\beq
P^h \to L^hX\to LX \label{g10}
\eeq
through some reduced principal subbundle $L^hX\subset LX$ with a
structure proper Lorentz group $L$. Thus, any Dirac spinor
structure is associated with a Lorentz reduced structure, but the
converse need not be true. There is the well-known topological
obstruction to the existence of a Dirac spinor structure
\cite{ger}. For instance, a Dirac spinor structure on a
non-compact manifold $X$ exists iff $X$ is parallelizable.

Hereafter, we restrict our consideration to Dirac spinor
structures on a non-compact (and, consequently, parallelizable)
world manifold $X$. In this case, all Dirac spinor structures are
isomorphic. Therefore, there is one-to-one correspondence
\be
z_h: P^h_{\rm s} \to L^hX\subset LX
\ee
between the Lorentz reduced structures $L^hX$ and the Dirac spinor
structures $(P^h_{\rm s}, z_h)$ which factorize through the
corresponding $L^hX$.  In particular, every Lorentz bundle atlas
$\Psi^h=\{z^h_\iota\}$ (\ref{lat}) of $L^hX$ gives rise to an
atlas
\mar{atl0}\beq
\ol\Psi^h=\{\ol z^h_\iota\}, \qquad z^h_\iota =z_h\circ \ol
z^h_\iota, \label{atl0}
\eeq
of a principal $\rL_{\rm s}$-bundle $P^h_{\rm s}$. We agree to
call $P^h_{\rm s}$ the spinor principal bundles.

Let $(P^h_{\rm s}, z_h)$ be the Dirac spinor structure associated
with a tetrad field $h$. Let
\mar{y1}\beq
S^h=(P^h_{\rm s}\times V)/\rL_{\rm s}\to X \label{y1}
\eeq
be a $P^h_{\rm s}$-associated spinor bundle whose typical fibre
$V$ carriers the spinor representation (\ref{b3213}) of a spinor
Lorentz group $\rL_{\rm s}$. One can think of sections of $S^h$
(\ref{y1}) as describing Dirac spinor fields in the presence of a
tetrad field $h$.

Let us consider an $L^hX$-associated bundle in Minkowski spaces
\be
M^hX=(L^hX\times M)/\rL=(P^h_{\rm s}\times M)/\rL_{\rm s}
\ee
and a $P^h_{\rm s}$-associated spinor bundle $S^h$ (\ref{y1}). It
is isomorphic to the cotangent bundle
\mar{int2}\beq
T^*X=(L^hX\times M)/\rL. \label{int2}
\eeq
Then, using the morphism (\ref{w01}), one can define a
representation
\mar{L4}\beq
\g_h: T^*X\ot S^h=(P^h_{\rm s}\times (M\ot V))/\rL_{\rm s}\to
 (P^h_{\rm s}\times \g(M\ot V))/\rL_{\rm s}=S^h \label{L4}
\eeq
of covectors to $X$ by Dirac $\g$-matrices on elements of a spinor
bundle $S^h$. Relative to a Lorentz bundle atlas $\{z^h_\iota\}$
of $LX$ and the corresponding atlas $\{\ol z_\iota\}$ (\ref{atl0})
of a spinor principal bundle $P^h_{\rm s}$, the representation
(\ref{L4}) reads
\be
y^A(\g_h(h^a(x) \ot v))=\g^{aA}{}_By^B(v), \qquad v\in S^h_x,
\ee
where $y^A$ are associated bundle coordinates on $S^h$, and $h^a$
are tetrad coframes. For brevity, we write
\mar{L4'}\beq
\wh h^a=\g_h(h^a)=\g^a,\qquad \wh
dx^\la=\g_h(dx^\la)=h^\la_a(x)\g^a. \label{L4'}
\eeq

Given the representation (\ref{L4'}), one can introduce a Dirac
operator on $S^h$ with respect to a principal connection on $P^h$.
Then, sections of a spinor bundle $S^h$ describe Dirac spinor
fields in the presence of a tetrad field $h$. Note that there is
one-to-one correspondence between the principal connections on
$P^h$ and those on a proper Lorentz bundle $L^hX=z_s(P^h)$. Then
it follows from Theorem \ref{gtg} that, since Lie algebras of
$GL_4$ and L obey the decomposition (\ref{g13}), any world
connection $\G$ yields a spinor connection $\G_s$ on $P^h$ and
$S^h$ \cite{book09,sard98a}.

This fact enables one to describe Dirac spinor fields in the
framework of metric-affine gravitation theory with general world
connections.

Dirac spinor fields in the presence of different tetrad field $h$
and $h'$ are described by sections of different spinor bundles
$S^h$ and $S^{h'}$. A problem is that, though reduced Lorentz
bundles $L^hX$ and $L^{h'}X$ are isomorphic, the associated
structures of bundles in Minkowski spaces $M^hX$ and $M^{h'}X$
(\ref{int2}) on the cotangent bundle $T^*X$ are non-equivalent
because of non-equivalent actions of a Lorentz group on a typical
fibre of $T^*X$ seen both as a typical fibre of $M^hX$ and that of
$M^{h'}X$. As a consequence, the representations $\g_h$ and
$\g_{h'}$ (\ref{L4}) for different tetrad fields $h$ and $h'$ are
non-equivalent \cite{sardz,sard98a}. Indeed, let
\be
 t^*=t_\m dx^\m=t_ah^a=t'_a{h'}^a
\ee
be an element of $T^*X$. Its representations $\g_h$ and $\g_{h'}$
(\ref{L4}) read
\be
\g_h(t^*)=t_a\g^a=t_\m h^\m_a\g^a, \qquad
\g_{h'}(t^*)=t'_a\g^a=t_\m {h'}^\m_a\g^a.
\ee
They are non-equivalent because no isomorphism $\Phi_{\rm s}$ of
$S^h$ onto $S^{h'}$ can obey the condition
\be
\g_{h'}(t^*)=\Phi_{\rm s} \g_h(t^*)\Phi_{\rm s}^{-1}, \qquad
t^*\in T^*X.
\ee

Since the representations (\ref{L4'}) for different tetrad fields
fail to be equivalent, one meets a problem of describing Dirac
spinor fields in the presence of different tetrad fields and under
general covariant transformations.

In order to solve this problem, let us consider a universal
two-fold covering $\wt{GL}_4$ of a group $GL_4$ and a
$\wt{GL}_4$-principal bundle $\wt{LX}\to X$ which is a two-fold
covering bundle  of a frame bundle $LX$ {\cite{law}. Then we have
a commutative diagram
\be
\begin{array}{ccc}
 \wt{LX} & \ar^\zeta & LX \\
 \put(0,-10){\vector(0,1){20}} &
& \put(0,-10){\vector(0,1){20}}  \\
P^h & \ar & L^hX
\end{array}
\ee
for any Dirac spinor structure (\ref{g10}) \cite{sard98a}. As a
consequence, $\wt{LX}/\rL_s=LX/L=\Si_{\rm T}$. Since $\wt{LX}\to
\Si_T$ is an $\rL_s$-principal bundle, one can consider an
associated spinor bundle $S\to \Si_T$ whose typical fibre is a
Dirac spinor space $V$. We agree to call it the universal spinor
bundle because, given a tetrad field $h$, the pull-back
$S^h=h^*S\to X$ of $S$ onto $X$ is a spinor bundle $S^h$ on $X$
which is associated with an $\rL_s$-principal bundle $P^h$. A
universal spinor bundle $S$ is endowed with bundle coordinates
$(x^\la, \si^\m_a,y^A)$, where $(x^\la, \si^\m_a)$ are bundle
coordinates on $\Si_T$ and $y^A$ are coordinates on a spinor space
$V$. A universal spinor bundle $S\to\Si_T$ is a subbundle of a
bundle in Clifford algebras which is generated by a bundle of
Minkowski spaces associated with an L-principal bundle
$LX\to\Si_T$. As a consequence, there is a representation
\mar{L7}\beq
\g_\Si: T^*X\op\ot_{\Si_T} S \to S, \qquad \g_\Si (dx^\la)
=\si^\la_a\g^a, \label{L7}
\eeq
whose restriction to a subbundle $S^h\subset S$ restarts the
representation (\ref{g11}).

Sections of a composite bundle
\mar{qqz}\beq
S\to \Si_{\rm T}\to X \label{qqz}
\eeq
describe Dirac spinor fields in the presence of different tetrad
fields as follows \cite{book09,sard98a}. Due to the splitting
(\ref{g13}), any world connection $\G$ on $X$ yields a connection
\mar{b3266}\ben
&& A_\Si = dx^\la\ot(\dr_\la - \frac14
(\eta^{kb}\si^a_\m-\eta^{ka}\si^b_\m)
 \si^\nu_k \G_\la{}^\m{}_\nu L_{ab}{}^A{}_By^B\dr_A) +
 \label{b3266}\\
&& \qquad d\si^\m_k\ot(\dr^k_\m + \frac14 (\eta^{kb}\si^a_\m
-\eta^{ka}\si^b_\m) L_{ab}{}^A{}_By^B\dr_A) \nonumber
\een
on the universal spinor bundle $S\to\Si_T$. Its restriction to
$S^h$ is the above mentioned spinor connection
\mar{b3212}\beq
\G_s=dx^\la\ot[\dr_\la +\frac14
(\eta^{kb}h^a_\m-\eta^{ka}h^b_\m)(\dr_\la h^\m_k - h^\nu_k
\G_\la{}^\m{}_\nu)L_{ab}{}^A{}_B y^B\dr_A] \label{b3212}
\eeq
defined by $\G$. The connection (\ref{b3266}) yields the so called
vertical covariant differential
\mar{7.10'}\beq
\wt D= dx^\la\ot[y^A_\la- \frac14(\eta^{kb}\si^a_\m
-\eta^{ka}\si^b_\m)(\si^\m_{\la k} -\si^\nu_k
\G_\la{}^\m{}_\nu)L_{ab}{}^A{}_By^B]\dr_A, \label{7.10'}
\eeq
on a fibre bundle $S\to X$ (\ref{qqz}). Its restriction to
$J^1S^h\subset J^1S$ recovers the familiar covariant differential
on the spinor bundle $S^h\to X$ relative to the spin connection
(\ref{b3212}). Combining (\ref{L7}) and (\ref{7.10'}) gives the
first order differential operator
\be
\cD=\si^\la_a\g^{aB}{}_A[y^A_\la- \frac14(\eta^{kb}\si^a_\m
-\eta^{ka}\si^b_\m)(\si^\m_{\la k} -\si^\nu_k
\G_\la{}^\m{}_\nu)L_{ab}{}^A{}_By^B],
\ee
on the fibre bundle $S\to X$ (\ref{qqz}). Its restriction to
$J^1S^h\subset J^1S$ is the familiar Dirac operator on a spinor
bundle $S^h$ in the presence of a background tetrad field $h$ and
a general linear connection $\G$.

Due to the decomposition (\ref{g13}), there is a canonical lift of
any vector field on $X$ onto bundles $P^h$ and $S^h$ though they
are not natural bundles \cite{godina,sard98a}. However, this lift,
called the Kosmann's Lie derivative, fails to be an infinitesimal
generator of general covariant transformations.

Since a world manifold $X$ is assumed to be parallelizable, a
universal spinor structure is unique, and a principal
$\wt{GL}_4$-bundle $\wt{LX}\to X$ as well as a linear frame bundle
$LX$ admits a functorial lift of any diffeomorphism $f$ of a base
$X$. This lift is defined by a commutative diagram
\be
\begin{array}{rcccl}
 &\wt{LX} & \ar^{\wt f_\rs} & \wt{LX}& \\
_{\wt z} &\put(0,10){\vector(0,-1){20}} & &
\put(0,10){\vector(0,-1){20}} &
_{\wt z} \\
& LX & \ar^{\wt f} & LX & \\
 &\put(0,10){\vector(0,-1){20}} &  & \put(0,10){\vector(0,-1){20}} & \\
& X & \ar^f & X  &
\end{array}
\ee
where $\wt f$ is the holonomic bundle automorphism of $LX$
(\ref{025}) induced by $f$. Consequently, a universal spinor
bundle $\wt{LX}$ is a natural bundle, and it admits the functorial
lift $\wt\tau_\rs$ of vector fields $\tau$ on its base $X$. These
lifts $\wt\tau_\rs$ are infinitesimal general covariant
transformations of $\wt{LX}$. Consequently, the composite bundle
$S\to X$ (\ref{qqz}) also is a natural bundle, and it possesses
infinitesimal general covariant transformations
\cite{book09,sard98a}.

\section{Appendix. Affine world connections}

The tangent bundle $TX$ of a world manifold $X$ as like as any
vector bundle possesses a natural structure of an affine bundle.
It is associated with a principal bundle $AX$ of oriented affine
frames in $TX$ whose structure group is a general affine group
$GA(4,\Bbb R)$. This structure group is always reducible to a
linear subgroup $GL_4$ since the quotient $GA(4,\Bbb R)/GL_4$ is a
vector space $\Bbb R^4$. The corresponding quotient bundle
$AX/GL_4$ is isomorphic to the tangent bundle $TX$. There is the
canonical injection of a frame bundle $LX\to AX$ onto a reduced
$GL_4$-principal subbundle of $AX$ which corresponds to the zero
section $\wh 0$ of $TX$.

Treating as an affine bundle, the tangent bundle $TX$ admits
affine connections
\mar{mos033}\beq
A= dx^\la\ot(\dr_\la + \G_\la{}^\al{}_\m(x) \dot x^\m\dot\dr_\al
+\si^\al_\la(x)\dot\dr_\al), \label{mos033}
\eeq
called affine world connections. They are associated with
principal connections on an affine frame bundle $AX$. Every affine
connection $\G$ (\ref{mos033}) on $TX$ yields a unique linear
connection
\mar{mos032}\beq
\G= dx^\la\ot(\dr_\la + \G_\la{}^\al{}_\m(x) \dot
x^\m\dot\dr_\al), \label{mos032}
\eeq
on $TX$. It is associated with a principal connection on a frame
bundle $LX\subset AX$. Conversely, by virtue of the well-known
theorem, any principal connection on a frame bundle $LX\subset AX$
gives rise to a principal connection on an affine frame bundle
$AX$, i.e., every linear connection on $TX$ can be seen as the
affine one. It follows that any affine connection $A$
(\ref{mos033}) on the tangent bundle $TX$ is represented by a sum
of the associated linear connection $\ol\G$ (\ref{mos032}) and a
soldering form
\mar{mos035}\beq
\si=\si^\al_\la(x) dx^\la\ot\dot\dr_\al \label{mos035}
\eeq
on $TX$, which is the $(1,1)$-tensor field
\mar{mos035'}\beq
\si=\si^\al_\la(x) dx^\la\ot\dr_\al \label{mos035'} \
\eeq
on $X$ due
to the canonical splitting (\ref{mos163}).

In particular, let us consider  the canonical soldering form
$\th_J$ (\ref{z117'}) on $TX$. Given an arbitrary world connection
$\G$ (\ref{B}) on $TX$, the corresponding affine connection
\mar{b1.97}\beq
 A=\G +\th_X, \qquad
A_\la^\m=\G_\la{}^\m{}_\nu \dot x^\nu +\dl^\m_\la, \label{b1.97}
\eeq
on $TX$ is a Cartan connection. Its torsion coincides with the
torsion $T$ (\ref{191}) of the world connection $\G$, while its
curvature is the sum $R+T$ of the curvature and the torsion of
$\G$.

There is a problem of a physical meaning of the tensor field $\si$
(\ref{mos035'}).

In the framework of so-called Poincar\'e gauge theory
\cite{heh,obukh}, it is treated as a non-holonomic frame, which is
a dynamic variable describing a gravitational field. This
treatment of $\si$ is wrong because a soldering form and a frame
are different mathematical objects. A frame is a (local) section
of a principal frame bundle $LX$, while a soldering form is a
global section of the $LX$-associated bundle
\be
\ol{LX}=(LX\times GL_4)/GL_4
\ee
whose typical fibre is a group $GL_4$, acting on itself by the
adjoint representation, but not left multiplications.

At the same time, a translation part of an affine connection on
$\Bbb R^3$ characterise an elastic distortion in gauge theory of
dislocations in continuous media \cite{kad,maly}. By analogy with
this gauge theory, a gauge model of hypothetic deformations of a
world manifold has been developed \cite{sard87,sardz}. They are
described by the translation part $\si$ (\ref{mos035'}) of affine
world connections on $X$ and, in particular, they are responsible
for the so called "fifth force" \cite{sard90,sardz}.

\end{document}